\documentclass{elsart}
\usepackage[english]{babel}
\usepackage{times}
\usepackage[iso]{umlaute}
\usepackage{epsfig}
\usepackage{setspace}
\usepackage{amsmath}

\begin{document} 
\begin{frontmatter}
\title{Combined Numerical and Experimental Investigation of Progressive Failure of Composites} 
\author[wittel]{H. Matthias Deuschle},
\author[wittel]{Falk K. Wittel\corauthref{cor}},\corauth[cor]{Corresponding author. Tel.: +49-711-685-7093; fax: +49-711-685-3706}
 \ead{wittel@isd.uni-stuttgart.de}
\author[ikp]{~Henry Gerhard},
\author[ikp]{Gerhard Busse},
\author[wittel]{Bernd-H. Kröplin}

\address[wittel]{Institute of Statics and Dynamics of Aerospace Structures (ISD), University of Stuttgart, Pfaffenwaldring 27, 70569 Stuttgart, Germany}
\address[ikp]{Institute for Polymer Testing and Polymer Science (IKP), University of Stuttgart, Pfaffenwaldring 32, 70569 Stuttgart, Germany}
\begin{abstract}
The combination of Finite Element Method (FEM) simulation and experimental photo-elasticity provides both qualitative and quantitative information about the stress field in a polymer composite and particularly along the fibre-matrix boundary. Investigations were made using model specimens containing up to five parallel glass fibres loaded at angles from $0^\circ$ to $15^\circ$. The material properties of the geometrically equivalent FE models are calibrated using the experimental photoelastic outputs. In contrast to the experimental results, FE simulations provide full 3D stress fields. For verification purposes, the 3D stress fields are reduced to two-dimensional synthetic photoelastic phase images, showing good agreement. Furthermore, detailed studies on the components of the stress tensor, particularly statements concerning the shear transmitted by fibre-matrix boundary, progressive fibre failure and the effect of load angle variation are presented.
\end{abstract}
\end{frontmatter}
\section{Introduction} \label{intro}
The progressive failure of long fibre composites, loaded in or at moderate angles to fibre direction, goes along with an increasing number of fibre fractures. The resulting load redistribution between fibre and matrix depends on the fibre and matrix properties and the behaviour of the fibre-matrix interface. The interface is mainly characterized by the transmitted shear which is itself strongly influenced by the load orientation \cite{Q1,Q2,Q3}. Assumptions for correct load redistribution from broken to unbroken fibres are of special interest for the realistic description of materials using models like fibre bundle models \cite{Q1,Q4}. Especially the form of the load sharing relation determines the functional behaviour of the model in the breakup process.

The breakup process can be observed experimentally with the non-destructive method of photoelasticity, using a confocal laser scanning microscope (LSM) \cite{In-situ-observation,improve-prec}. Complex stress states are visualized by transillumination leading to 2D phase images clearly showing lines of equal difference between principal stresses (isochromates) and their gradients \cite{cloud}. Unlike standard microscopes, the LSM provides information within any particular volume element of a transparent specimen while other information is effectively removed by a spatial filter \cite{klish-etal}. Unfortunately phase images are difficult to interpret, since in situ observation neither allows investigations on the stress variation in direction of specimen thickness nor on the composition of the stress tensor. To overcome these difficulties, comparative FE simulations are accomplished and calibrated to the experimental output. To obtain a common basis of comparison, the 3D results of the simulation have to be transferred into synthetic 2D phase images. This post processing reproduces the physical transillumination according to the experimental photoelasticity. The combined numerical and experimental approach allows a reliable calibration of the FE model, providing verified 3D stress fields as a basis for the presented studies.

The paper is organized as follows: After a short description of the experimental procedure in section 2, section 3 addresses the FE model and simulation. Section 4 covers the calibration and the interpretation of the experimental results. By means of the verified FE simulation, section 5 presents the composition of the stress tensor existing in the matrix before and after fibre fracture and its effect on the neighbouring fibre. Further studies focus on the effect of fibre-to-load orientation on both fibre and matrix loading and particularly on the shear component of the stress tensor along the fibre-matrix interface. Finally the progressive fracture behaviour at different load angles is investigated, followed by a discussion in section 6.
\section{Experimental photoelasticity with model samples}\label{experiment}
\begin{figure}[htb] 
  \begin{center}
    \epsfig{bbllx=19,bblly=19,bburx=576,bbury=375,file=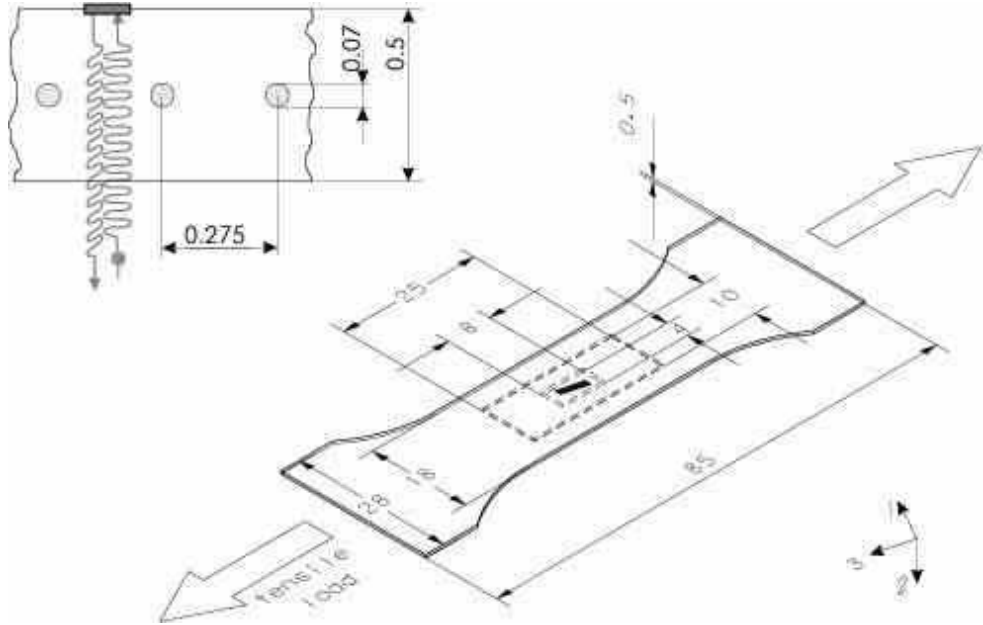, width=13cm} 
    \caption{Geometric situation of a five-fibre specimen ($15^\circ$), including the two sections covered by simulation (dashed) and light path for reflection mode}    \label{fig:1}
  \end{center}
\end{figure}

Photoelasticity is a non-destructive method of observing preferably 2D stress distribution in a specimen \cite{Grundl-Sp-Opt}. For that purpose a photoelastically active medium is required, such as polycarbonate Makrofol DE 1-1 (by Bayer \cite{bayer}) used as matrix material. When transilluminating the specimen (see fig. \ref{fig:1}), the linear polarized light is separated into two rectangular components according to the existing principal stress directions. Due to loading, the photoelastically active medium changes its refraction index, resulting in a phase shift between the two components of light, which have passed through the specimen at different velocities. Superposing both components, erasement respectively duplication of amplitude is the consequence. The resulting projection of the specimen is a grey scale image including isochromates (lines of equal principal stress differences) and isoclines (lines of equal principal stress directions). Our experimental setup uses circular polarized light which is indifferent against direction. Therefore no isoclines appear. Further details on the experimental procedure can be found in \cite{In-situ-observation,improve-prec}. The description of the following transformation into photoelastic phase images is published in \cite{improve-prec}. The resulting images clearly indicate the darkest and lightest isochromates by a black-to-white phase jump containing information about the direction of the gradient. If the specimen's thickness $d$ and the order of a phase shift $n$ at a single point of the image is known, the existing principal stress difference $\sigma_1-\sigma_2$ can be calculated for the whole image using the photoelastic constant $S$ according to 
\begin{equation}\label{equ:2000}
\sigma_1-\sigma_2=S \cdot n/d.
\end{equation}
$S$ characterizes the photoelastic activity. Since the laser beam passes the specimen in reflection mode twice, $d$ in Eq.\ref{equ:2000} is twice the specimen thickness. Note that $S$ behaves non-linearly for increasing load approximated by the experimentally determined function $S=f(\sigma_1-\sigma_2)$ \cite{Grundl-Sp-Opt}. However, more accurately treated $S=f(\sigma_1,\sigma_2)$ is a function of $\sigma_1$ and $\sigma_2$. The final experimental output is a 2D field of integrated, scalar principle stress differences which means that neither information on stress variations in specimen thickness nor components of stress are available. As the experiment deals with a 3D state of stress, the result does not reproduce the actual principal stress difference. It only accounts for those components existing in the plane rectangular to the way of light. For clarity, the result value is called secondary principle stress difference $\Delta\sigma_{sps}$.

Both the experimental and numerical program contain pure matrix specimens to define the photoelastic constant $S$, single fibre specimens (fibre diameter $\approx 70-80\mu m$) and specimens containing 5 parallel fibres oriented $0^\circ, 5^\circ, 10^\circ$ and $15^\circ$ towards the direction of load. Note that all stress indices throughout this work refer to fig. \ref{fig:1} with the triad aligned with the fibre orientation.
\section{The FE-simulation}\label{fe-sim}
\begin{figure}[htb] 
  \begin{center}
    \epsfig{bbllx=19,bblly=19,bburx=576,bbury=332,file=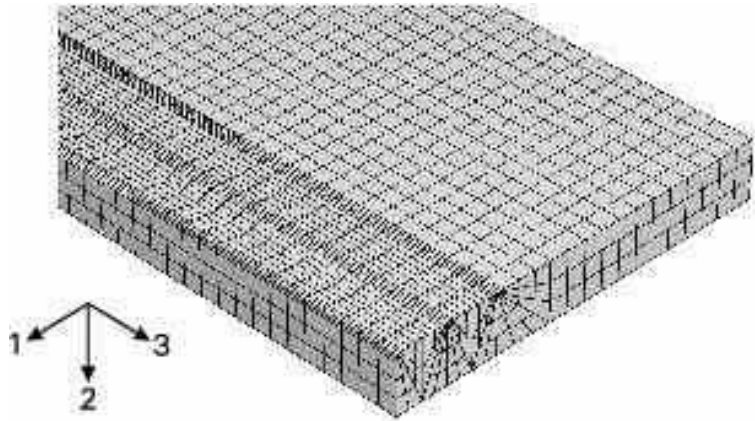, width=10cm} 
    \caption{Quarter-model of inner section with 8-node thermally coupled brick elements with trilinear displacement and temperature using reduced integration}    \label{fig:2}
  \end{center}
\end{figure}

The model geometry corresponds to the used specimen in fibre length, diameter and distance, matrix thickness and positions of fibre fractures. The overall dimensions of the model were chosen smaller than those of the specimen, but large enough to observe far field stress. Fig. \ref{fig:1} includes the model dimensions in comparison to the specimen size $[mm]$. To save computation time, the model is divided into an outer, roughly discretized and an inner, finer discretized section embedding the fibres. The inner section (quarter-model) of a meshed five-fibre model ($0^\circ$) including the cut-out for the embedded fibres is magnified in fig. \ref{fig:2}. The irregularities of the mesh in the fibre plane are due to the demand, that the neighbouring matrix node has to be exactly in the plane of fracture to obtain comparable results at each fracture position.

Elastic-plastic material behaviour is used for the matrix description combined with linear-elastic fibres. Since the simulation covers the cooling-down process of the specimen during preparation, the material definition also contains linear thermal expansion coefficients for both materials.
\section{Combined numerical and experimental approach}\label{Zusammenf}
\subsection{Synthetic phase images and calibration}\label{calibration}
To obtain reliable FEM results, the material properties were calibrated using the experimental photoelastic phase images. To obtain a common basis of comparison, the FEM result is integrated in the direction of the model thickness. This post processing only accounts for the secondary principal stress difference $\Delta\sigma_{sps}$, which is the photoelastically relevant component of the stress tensor  (see sec.\ref{experiment}). At each node $\Delta\sigma_{sps}$ is obtained by a simple 2D principal axis transformation of the involved components of stress according to
\begin{equation}\label{equ:1000}
\Delta\sigma_{sps}=\sigma_1-\sigma_2=\sqrt{(\sigma_{11}-\sigma_{33})^2+4\sigma_{13}^2}.
\end{equation}
Due to the non-linear behaviour of the photoelastic activity (see sec.\ref{experiment}), $\Delta\sigma_{sps}$ has to be transformed into the corresponding order of phase shift $n$ before integration. This value is calculated and normalized to unit thickness ($d=1$) for each node using Eq. (\ref{equ:2000}). Its weighted sums result in a 2D field of orders of phase shift $n$, which is visualized by a synthetic phase image \cite{studi}. These images are used for visual matching with the experimental phase images to determine the calibrated FEM material properties like the constitutive behaviour of the fibre material and its thermal expansion coefficient. Note that literature values for the {\sc Poisson} ratio of the fibre and matrix material ($\nu^f=0.21$, $\nu^m=0.39$) are used.

\begin{figure}[htb] 
  \begin{center}
    \epsfig{bbllx=19,bblly=19,bburx=576,bbury=257,file=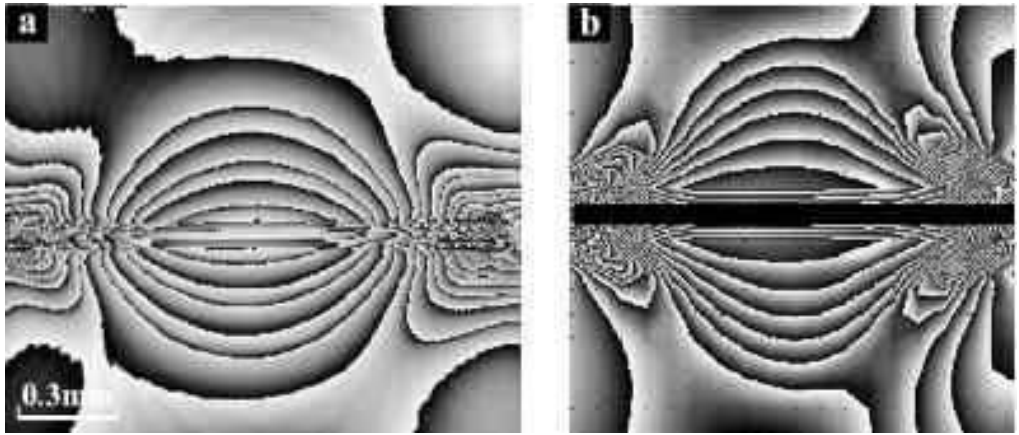, width=13.5cm} 
    \caption{Comparison of experimental $(a)$ and synthetic $(b)$ phase images of single-fibre specimen at a an applied tensile load of 54 MPa}  \label{fig:3}
  \end{center}
\end{figure}

The experimentally measured constitutive behaviour of the matrix was used for the FE simulations. Based on the implemented matrix behaviour, the fibre stiffness is calibrated matching the number of phase jumps of the experimental (fig. \ref{fig:3}a) and synthetic (fig. \ref{fig:3}b) phase images. Note that the relieve of the matrix by the embedded fibre leads to phase jumps surrounding the central part of the fibre.
The linear thermal expansion coefficient of the matrix is calculated by the shrinkage of the specimen in direction of thickness during cooling-down. Furthermore, the resulting residual thermal stresses are used for the adaptation of the linear thermal expansion coefficient of the fibre.
\subsection{Interpreting the photoelastic results}\label{resultsinterp}
\begin{figure}[htb] 
  \begin{center}
    \epsfig{bbllx=19,bblly=19,bburx=576,bbury=570,file=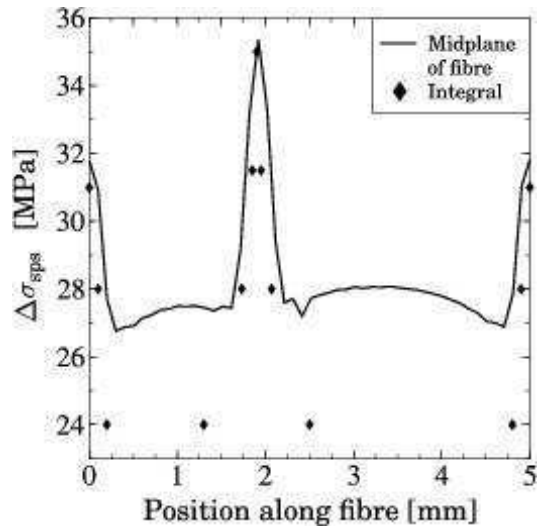, width=7cm} 
    \caption{Comparison of $\Delta\sigma_{sps}$ for a single-fibre specimen containing one fracture}  \label{fig:4}
  \end{center}
\end{figure}

\begin{figure}[htb] 
  \begin{center}
    \epsfig{bbllx=19,bblly=19,bburx=576,bbury=308,file=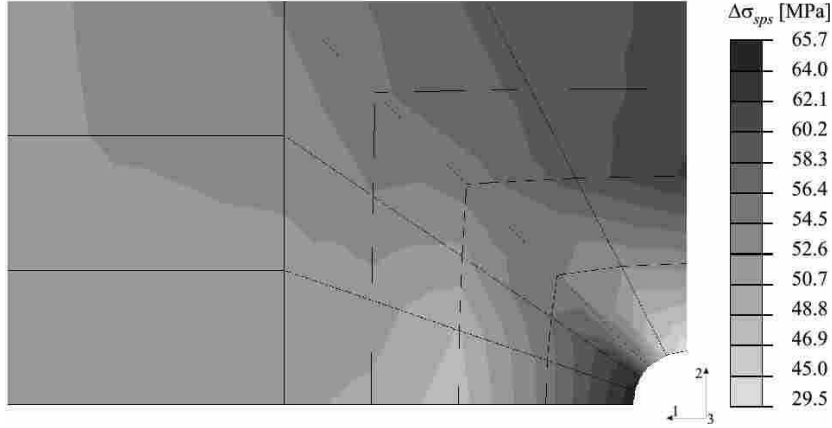, width=11.0cm} 
    \caption{$\Delta\sigma_{sps}$ in plane of fibre break in a $0^\circ$ single-fibre specimen at an applied tensile load of 56 MPa}  \label{fig:5}
  \end{center}
\end{figure}

A direct advantage of the combined approach is the possibility to estimate the significance of the experimental photoelastic results. For example the FE result allows a comparison of experimental integral results with the actual local stress states. Fig. \ref{fig:4} deals with a $0^\circ$ single-fibre specimen after one fibre fracture located at $1.9$mm with an applied tensile load of $27$ MPa. The difference between the integral and the mid-plane value of $\Delta\sigma_{sps}$ is an indirect measure for the homogeneity of the loading over specimen thickness and proves the necessity for 3D treatment. As this difference tends towards zero at the position of the fibre fracture, the increased matrix loading due to the fibre break extends over the whole specimen thickness. The large difference at less loaded positions is an evidence for a higher fluctuation of $\Delta\sigma_{sps}$ along the light path.
The definition of an algorithm to correct the deviations remains difficult because of the field of the photoelastically relevant $\Delta\sigma_{sps}$. In opposite to the actual rotational symmetric state of stress around the fibre, $\Delta\sigma_{sps}$ is not axisymmetric (see fig. \ref{fig:5}). An additional challenge for formulating a correction algorithm is the necessary decomposition of the scalar photoelastic result.
\section{Results}\label{results}
The presented FE results are based on the calibrated single fibre model with respect to material properties. This model is used for the fundamental characterization of the fibre-matrix interaction by showing the composition of the secondary principle stress difference (see sec.5.1). In the following, the problem is extended to a five-fibre model for studies on load redistribution due to fibre fracture and interaction of neighbouring fibres (see sec.5.2). Furthermore, results on the influence of fibre-to-load angle alteration on the fibre loading (see sec.5.3.1), the loading of matrix (see sec.5.3.2) and the progressive fracture behaviour (see sec.5.3.3) are presented.
\subsection{Composition of the secondary principal stress difference}\label{composition}
\begin{figure}[htb] 
  \begin{center}
    \epsfig{bbllx=19,bblly=19,bburx=576,bbury=566,file=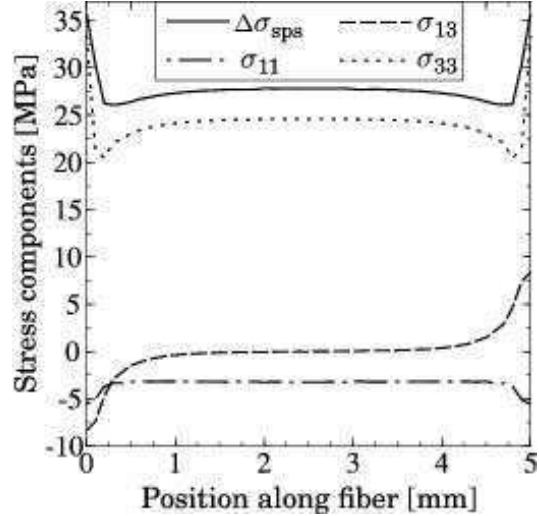, width=7cm} 
    \caption{Components of $\Delta\sigma_{sps}$ in the matrix along fibre-matrix interface in a $0^\circ$ single-fibre specimen without fracture at an applied tensile load of 25 MPa}  \label{fig:6}
  \end{center}
\end{figure}

Due to the calculated stress tensor the contribution of the different stress components to the photoelastically visualized $\Delta\sigma_{sps}$ is known (see fig. \ref{fig:6}).
It is clearly visible that the main component is the stress in fibre direction $\sigma_{33}$. The absolute value of the shear stress $\sigma_{13}$ has its maxima at both ends of the fibre, due to the load transmission from matrix to fibre. The cooling-down process of the specimen during preparation leads to a compressive stress component $\sigma_{11}$ around the fibre. Due to the rather complex composition of $\Delta\sigma_{sps}$ even for single-fibre specimens aligned with the load direction, the interpretation of the scalar experimental output seems to be difficult.
\subsection{Load redistribution due to fibre fracture}\label{resultslrdist}
\begin{figure}[htb] 
  \begin{center}
    \epsfig{bbllx=19,bblly=19,bburx=576,bbury=306,file=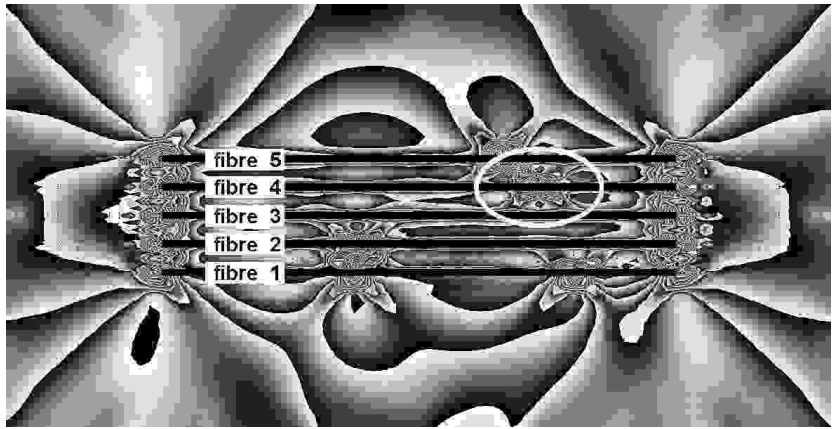, width=11cm} 
    \caption{Synthetic phase image of 5-fibre $0^\circ$ specimen after 5 fibre fractures at an applied tensile load of 45 MPa}  \label{fig:7}
  \end{center}
\end{figure}

Load redistribution due to a fibre fracture leads to an increased matrix loading in a butterfly-shaped area around the fracture (see fig. \ref{fig:12}a and earlier photoelastic investigations \cite{In-situ-observation}). This influences both, the state of stress of the matrix along the neighbouring fibre-matrix interface and the fracture behaviour of the neighbouring fibre itself. It is reflected in a remarkably high probability of the fibre to break in the sphere of influence of a neighbouring fracture. Two pairs of consecutive fractures are shown in fig. \ref{fig:7}. The FE simulation proves the butterfly shape of $\Delta\sigma_{sps}$. 

\begin{figure}[htb] 
  \begin{center}
    \epsfig{bbllx=19,bblly=19,bburx=576,bbury=279,file=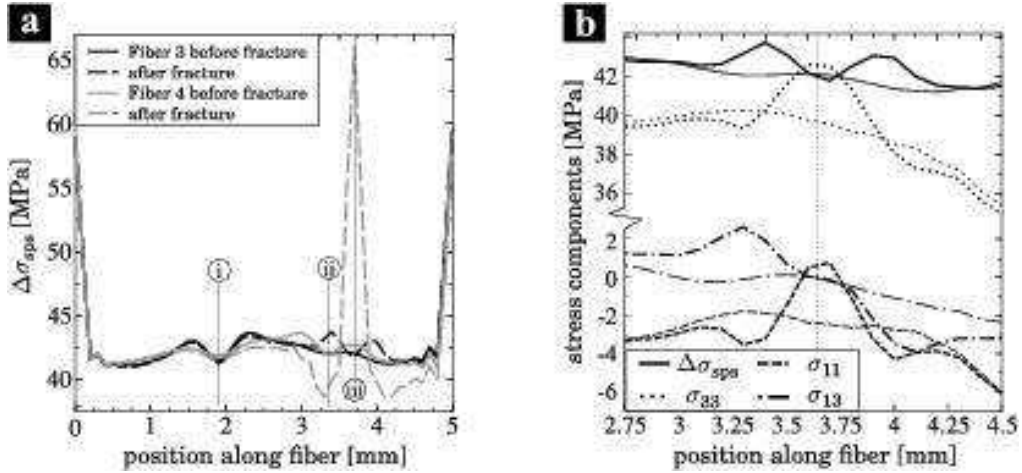, width=13.5cm} 
    \caption{a: $\Delta\sigma_{sps}$ in matrix before and after fracture of fibre 4; b: Components of $\Delta\sigma_{sps}$ at the neighbouring fibre-matrix interface 3 before and after fibre fracture (thin/bold lines)}  \label{fig:8}
  \end{center}
\end{figure}

For a quantitative description of the load redistribution, $\Delta\sigma_{sps}$ and its components are extracted along fibre-matrix interfaces (see fig.\ref{fig:8}). Note that fibre labels refer to fig. \ref{fig:7}.
Looking at the direct surrounding of the breaking of fibre 4, the fracture causes a maximum of $\Delta\sigma_{sps}$ in the matrix (see fig. \ref{fig:8}a) consisting mainly of two stress components: stress in fibre direction $\sigma_{33}$ and shear stress $\sigma_{13}$. The increase of $\sigma_{33}$ is caused by the complete tensile load redistribution from fibre to matrix, while $\sigma_{13}$ increases because of two newly emerged areas of load transmission from matrix to fibre.

The consequences on the neighbouring fibre 3 are shown in fig. \ref{fig:8}b where the fracture leads to a minimum of $\Delta\sigma_{sps}$ in the fracture plane with two surrounding maxima. This typical shape is a cut through the butterfly area. Furthermore, fig. \ref{fig:8}a includes the decreasing influence of the existing fracture in fibre 2  on the loading along fibre 3 and 4 (i) and the influence of the fracture of fibre 5 on fibre 4 (ii) before it consecutively breaks at the increased local maximum of $\Delta\sigma_{sps}$ (iii). Looking at the composition of $\Delta\sigma_{sps}$ in an enlarged zone along the neighbouring fibre 3 (see fig. \ref{fig:8}b), $\sigma_{33}$ increases in the fracture plane analogously to the direct fracture surroundings in fibre 4. At the same position, the absolute value of compressive stress $\sigma_{11}$ decreases to about $0$ MPa, which might weaken the fibre-matrix bonding. The maximum of the absolute value of $\sigma_{13}$ causes the two local maxima of $\Delta\sigma_{sps}$, which agree with the experimentally observed cascade positions. The butterfly shape and the probability of the neighbouring fibre to break in typical cascade formation (see fig. \ref{fig:7}, fibres 5 and 4) therefore arise from the behaviour of the shear stress $\sigma_{13}$. The comparable relevance of $\sigma_{33}$ for the damage evolution is addressed in sec.\ref{faserorient}.
\subsection{Influence of fibre orientation}\label{resultsorientf}
To study the progressive failure of long fibre composites under shear stress loading, the test series contains several 5-fibre specimens at an increasing load angle. By means of the presented simulations we can extract the states of stress during the experiment, giving rise to interpretation of the gradual failure.
\subsubsection{Influence of fibre orientation on the loading of fibre}\label{faserbel}
Since photoelastic experiments show the effect of the embedded fibre but not the stress state of the fibre itself, the investigation relies on the results of FE simulations. The presented study focuses on the tensile stress $\sigma_{33}$, which is assumed to be one reason for fibre failure.
\begin{figure}[htb] 
  \begin{center}
    \epsfig{bbllx=19,bblly=19,bburx=576,bbury=564,file=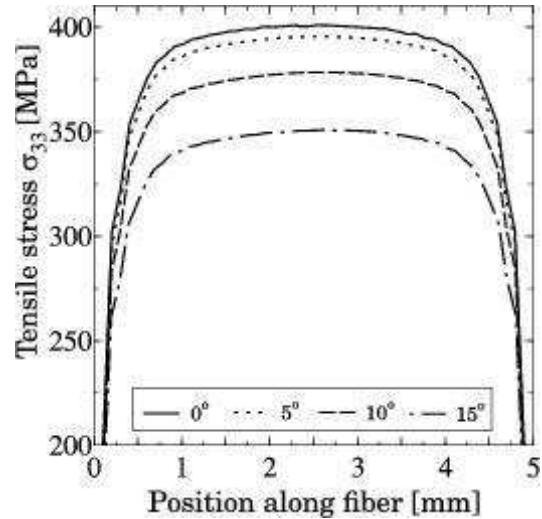, width=7cm} 
    \caption{Tensile stresses in the outer fibres of the five-fibre package at an applied load of $30$ MPa}  \label{fig:9}
  \end{center}
\end{figure}
$\sigma_{33}$ increases rapidly in the areas of load transmission up to a relatively constant value over the mid-fibre positions. Looking at the influence of variation of fibre orientation, the expected decrease of load absorption by the fibres with increasing load angle is observed and in agreement with experiments. Additionally the numerical results prove the experimental observation, that outer fibre tends to break first. The tensile stress in the outer fibre of a package is about $5\%$ higher than in the middle fibre.
 \subsubsection{Influence of fibre orientation on the loading of matrix}\label{resultsorientm}
\begin{figure}[htb] 
  \begin{center}
    \epsfig{bbllx=19,bblly=19,bburx=576,bbury=306,file=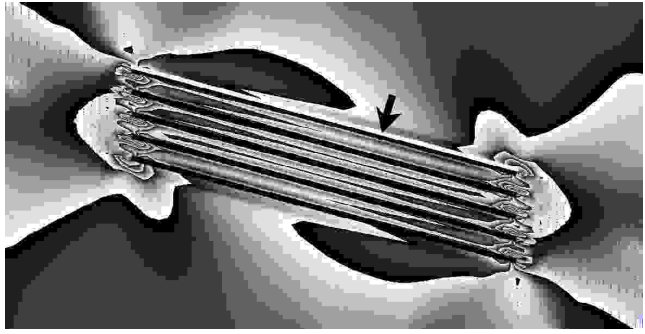, width=8.5cm} 
    \caption{Synthetic phase image of a five-fibre model at $15^\circ$ load angle}  \label{fig:10}
  \end{center}
\end{figure}

\begin{figure}[htb] 
  \begin{center}
    \epsfig{bbllx=19,bblly=19,bburx=576,bbury=563,file=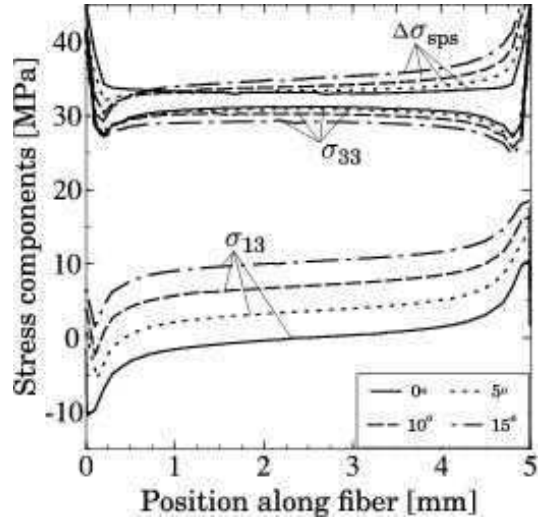, width=7cm} 
    \caption{Components of $\Delta\sigma_{sps}$ along the path of fig. \ref{fig:10} at $34$ MPa}  \label{fig:11}
  \end{center}
\end{figure}

To understand the influence of load angle alteration on the stress distribution existing in the matrix, $\Delta\sigma_{sps}$ is decomposed with special interest in $\sigma_{13}$. Therefore the stresses along the fibre-matrix interface are chosen for further evaluation (see figs. \ref{fig:10},  \ref{fig:11}).
At higher load angles, the decreasing load absorption by the fibres results in a moderate increase of $\Delta\sigma_{sps}$ (see fig. \ref{fig:11}). Both the synthetic phase images (see fig. \ref{fig:10}) and the extracted numerical output (see fig. \ref{fig:11}) show a maximum of $\Delta\sigma_{sps}$ in the area of load transmission at the right end of the fibre. Additionally, the maximal relief of the matrix by the embedded fibre is shifted to the left. Looking at the components of $\Delta\sigma_{sps}$ the stress in fibre direction $\sigma_{33}$ decreases slightly due to the rotation of coordinates but shows little asymmetric behaviour. In return  $\sigma_{11}$ is shifted to  a tensile stress (see for $15^\circ$ fig. \ref{fig:13}a]. The largest sensitivity against the load angle is found for the shear stress $\sigma_{13}$. Loaded at an angle, the shear stress is a superposition of two components: The first one arises from the above-described process of load transmission from the matrix to the fibre (see sec.\ref{composition}) leading to the extreme values at the fibre ends. The second component is constant, which originates from the rotation of the fibre-matrix interface and shifts the shear stress $\sigma_{13}$ up to higher values. As a result, the inflection point in mid-fibre position is no longer at zero but describes the amount of the second component of $\sigma_{13}$.
\subsubsection{Influence of fibre orientation on the fracture behaviour}\label{faserorient}
\begin{figure}[htb] 
  \begin{center}
    \epsfig{bbllx=19,bblly=19,bburx=576,bbury=276,file=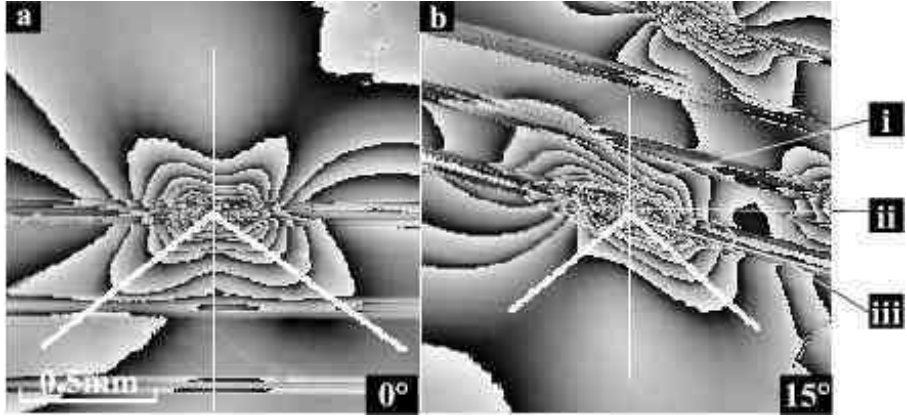, width=12cm} 
    \caption{Photoelastic phase images of fracture areas. Thin lines define the fracture plane, while bold lines connect the points of largest $\Delta\sigma_{sps}$ at increasing distance from the fracture}  \label{fig:12}
  \end{center}
\end{figure}

\begin{figure}[htb] 
  \begin{center}
    \epsfig{bbllx=19,bblly=19,bburx=576,bbury=266,file=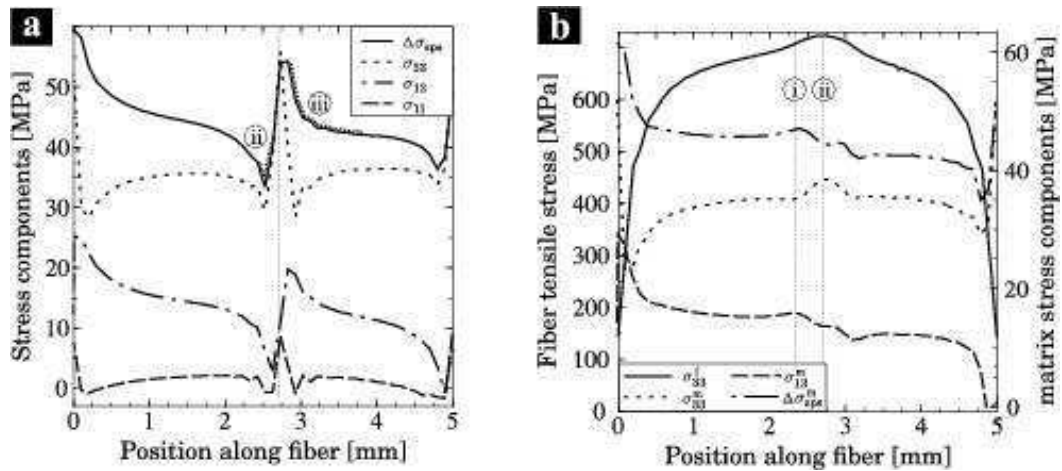, width=14cm} 
    \caption{a: Components of $\Delta\sigma_{sps}$ after fibre fracture along the lowest fibre-matrix interface of fig.\ref{fig:12}b at an applied load of $44$ MPa and $15^\circ$ b: Loading of the neighbouring fibre and the matrix along its interface}  \label{fig:13}
  \end{center}
\end{figure}

In order to understand the progressive failure of multi-fibre specimen, one needs to understand the influence of fibre orientation on the loading of fibre and matrix. First we focus on the direct fracture zone (see fig.\ref{fig:12}b$^i$), where those areas of the fibre-matrix interface, which form an acute angle with the fracture plane are higher loaded (see figs.\ref{fig:12}b$^{iii}$,\ref{fig:13}a$^{iii}$). The similar effect was previously observed for the fibre ends (see sec.\ref{resultsorientm}). Along the opposite interface a distinctive minimum of $\Delta\sigma_{sps}$ is observed (see figs.\ref{fig:12}b$^{ii}$, \ref{fig:13}a$^{ii}$).
The consequences of different fibre orientations can be seen in the numerically determined stress state (see fig.\ref{fig:13}a). As shown in fig.\ref{fig:11}, $\sigma_{13}$ is again affected by the variation of fibre orientation and forms the crucial factor for the described minimum of $\Delta\sigma_{sps}$  (see figs.\ref{fig:12}b$^{ii}$, \ref{fig:13}a$^{ii}$). The numeric results also show that $\sigma_{11}$ and $\sigma_{33}$ are almost insensitive with respect to the load angle, as followed from their symmetry to the fracture plane (see fig.\ref{fig:13}a), and behave similarly to $0^\circ$ specimens. Note that $\sigma_{11}$ is simply shifted to positive values due to a superposed tensile stress component exceeding the compressive residual stress around the fibre (see fig.\ref{fig:13}a).

Analyzing the surrounding matrix area, we observe that the lines of maximum $\Delta\sigma_{sps}$ do not rotate according to the fibre orientation (compare figs.\ref{fig:12}a and \ref{fig:12}b), so the orientation of the `butterfly wings' is not affected by the fibre orientation. In contrast, the value of $\Delta\sigma_{sps}$ along the bold line is subject to the variation of fibre orientation, in such a way that the `butterfly wings' are distorted asymmetrically.

Numerous experimental phase images including earlier investigations \cite{improve-prec,klish-etal} reveal two different types of progressive fracture events: cascade-like propagation of fibre fractures (see fig.\ref{fig:7} fibres 4/5) and those in fracture plane (see fig.\ref{fig:7} fibres 1/2). The differences in the stress composition at the two predetermined fracture positions (fig.\ref{fig:13}b:i, ii) are used for interpretation. The cascade-like propagation (see fig.\ref{fig:13}b$^i$) is not due to the elevation of the tensile stress of the fibre $\sigma_{33}^f$ but might be the result of the influence of the shear stress component $\sigma_{13}^m$ at the fibre-matrix interface. On the other hand the maximum of $\sigma_{33}^f$ is the crucial factor for the second type of progressive fracture events and coincides with the maximum of $\sigma_{33}^m$ in plane of fracture (see fig.\ref{fig:13}b$^{ii}$).
 
Including the influence of load angle variation, the neighbouring fibre has jsut one predestined cascade position of break, whereas at $0^\circ$ the undistorted butterfly has two equally probable fracture positions per neighbouring fibre. Independent of the load angle, the fracture position is located on the bold line (see fig.\ref{fig:12}b).
\section{Discussion}\label{discussion}
As a basis for this work we have documented and evaluated the influence of the load direction on the progressive failure of multi-fibre specimens via confocal laser scanning photoelasticity (see fig.\ref{fig:12}). The following calibration of the FE model on simple specimens containing one fibre gives the necessary material properties for the more sophisticated simulations (see sec.\ref{calibration}). Extensive agreement of experimental and synthetic phase images allows reliable quantitative statements on the load redistribution during the failure process. In addition to the experimental photoelasticity, the FEM simulation provides a stress tensor used for further studies (see sec.\ref{composition}). In principle the composition of the photoelastically visualized secondary principle stress differences strongly depends on the applied tensile load and load orientation (see fig.\ref{fig:11}). We have shown that the shear stress component is an important factor e.g. for the butterfly shape of increased matrix loading due to fibre fracture. The significance of the shear stress increases for higher load angles (see sec.\ref{resultsorientm}).

The FEM simulation includes the photoelastically inactive fibres and proves the expected reduction of load absorption by the fibres for increasing load angle (see sec.\ref{faserbel}). Earlier photoelastic investigations and the presented simulation results reveal two different types of progressive fracture events, one influenced by the shear stress component (cascade-like failure) and the other one by the tension in fibre direction (see sec.\ref{faserorient}). Since the butterfly is distorted for angular load, only one position at the neighbouring fibre becomes the predestined cascade-like fracture position. Note that the orientation of the cascade is independent of the fibre direction.

Despite of the simplifications and assumptions made in the model, the presented simulations contribute to the comprehension of load redistribution processes, the load angle influence, the fibre interaction and the fracture propagation. Particularly due to the iterative calibration process, it was necessary to neglect creep rates and adaptive fibre-matrix debonding. The presented results show, that the interpretation of photoelastic phase images visualizing complex states of stress has to be broadened by a combined numerical investigation to gain full insight. Furthermore, the presented numerical approach can be extended beyond the physical limits of photo elasticity such as small fiber distances or complex 3D fibre arrangements.
\section{Acknowledgment} 
The presented work is partly funded by the German Science Foundation (DFG) within the Collaborative Research Center SFB 381 'Characterization of Damage Development in Composite Materials using Non-Destructive Test Methods' under projects B1/C7.

\end{document}